\documentclass[smallcondensed]{svjour3}    

\smartqed  

\usepackage{graphicx}
\usepackage{mathptmx}      

\usepackage{latexsym}

\begin{document}

\title{Quantification of tidal parameters from Solar system data}


\author{Val\'ery Lainey 
}


\institute{V. Lainey \at
              IMCCE, Observatoire de Paris, PSL Research University, CNRS-UMR8028 du CNRS,
UPMC, Lille-1, 77 Av. Denfert-Rochereau, 75014, Paris, France \\
              Tel.: +33-1-40512269\\
              Fax: +33-1-46332834\\
              \email{lainey@imcce.fr} }        
\maketitle

\begin{abstract}
Tidal dissipation is the main driver of orbital evolution of natural satellites and a key point to understand the exoplanetary system configurations.
Despite its importance, its quantification from observations still remains difficult for most objects of our own Solar system. In this work, we overview the method
that has been used to determine, directly from observations, the tidal parameters, with emphasis on the Love number $k_2$ and the tidal quality factor $Q$. Up-to-date values of these tidal parameters are summarized. 
Last, an assessment on the possible determination of the tidal ratio $k_2/Q$ of Uranus and Neptune is done. This may be particularly relevant for coming astrometric campaigns 
and future space missions focused on these systems.


\end{abstract}

\section{Introduction}\label{intro}

While fundamental for our understanding of the Solar system and the long-term evolution of the exo-planetary systems, tidal parameters are difficult to determine from observations. Indeed, they provide a small dynamical signal, both on natural and artificial celestial objects, in comparison to other perturbations arising in the system (see next section). However, tides do provide large effects 
on the long term evolution of the natural bodies. This is why estimations of the tidal parameters have often been done from the assumed past evolution of planets and moons.
In that respect, estimation of the average tidal ratio $k_2/Q$ for the giant planets was done already back in the 60s by Goldreich \& Soter (1966). We recall that $k_2$ is the first of the second 
order Love number (Love 1911), while $Q$ is a quality factor relative to the amount of mechanical energy dissipated by friction inside an object (Goldreich \& Soter, 1966). Both are unit-less 
quantities. Considering the average secular drift over 4.5 Byr on the semi-major axes of the innermost main moon of the Jupiter, Saturn and Uranus systems, the authors obtained an 
averaged estimation of $k_2/Q$. While having been the reference for several decades, these values were estimated assuming a specific formation and evolution scenario. In particular, 
alternative formation models now exist for these moons, that may suggest younger age for some of the main moons (Crida \& Charnoz (2012), \'Cuk (2014)). Moreover, resonances crossings 
and possibility of strong dissipative episodes of the moons makes the question even more complex.

Another way for constraining the tidal ratio $k_2/Q$ may be, at least for very active bodies, to simply consider the amount of heat radiated from their surface, and after having considered the 
presence of radiogenic heating. IR-emission values may be used for Io (Matson et al. 1981) and Enceladus (Spencer et al. 2006) to this task. However, this line of reasoning assumes a 
thermal equilibrium state to infer the amount of heat produce in the whole interior, which is not granted.

As a result, a direct measurement of tidal parameters done from observations sounds more reliable. This can be assessed by monitoring the motion of celestial objects evolving in the vicinity 
of the tide raised body. To that aim, astrometric data associated with artificial satellites, spacecraft and natural moons motions are relevant. Here, astrometry has to be understood in its 
general sense of any measurement able to provide information on position or velocity of celestial objects into space. Besides classical imaging astrometry using star background, it includes 
radiometric, laser ranging, VLBI, GNSS and even sometimes photometric measurements.

In the next section, we briefly recall the method that is used for determining tidal parameters from astrometry. In section \ref{sec:Tidal} we summarize the most up-to-date values of the 
major tidal parameters we know from astrometry, so far. In Section \ref{sec:IcyG}, we study the feasibility of determining, from observations, the tidal ratio $k_2/Q$ for the two icy giant 
planets of our Solar system.

\section{Methodology}\label{methodo}

The methods used to constrain tidal parameters from astrometric data of the natural objects of the Solar system on one hand, and of the spacecraft on the other hand, are actually extremely similar. They require three steps: i) the development of an orbital model of the objects studied; ii) the gathering of observation sets of the objects; iii) the fitting of the dynamical model to the observations. 

The modeling of the system stands on a $N$-body code that takes into account all the perturbations that may influence the orbit, at the level of accuracy of the astrometric observations. 
In the case of space geodesy, only one body is integrated over time, making $N=1$. Still, several other celestial objects have to be considered as perturbers of the spacecraft/artificial 
satellite dynamics. For the perturbations, the minimum is to consider all objects as point mass, but it is most of the time necessary to take into account the harmonic expansion of the 
gravity fields. In that case, the direction of the north pole and prime meridian into space of each body needs to be considered, generally by mean of forced frequencies (Archinal et al. 2011), and ultimately fitted. Depending on the system 
studied, the list of perturbations that have to be introduced can be pretty long. Let us just mention here, the extended gravity fields of the objects, the object's precession and nutations, 
the forced librations on rotation, the tidal effects, the relativistic effects. In the case of space geodesy, we may add to the perturbations' list, the drag into the atmosphere (if any) of the primary, 
the solar and planetary radiation pressure, the wheel-off loadings, etc. Once a proper modeling is set, and using Newton's second law, one will have to integrate the ordinary differential equation 
(ODE) of second order
\begin{eqnarray}
\frac{d^2{\bf r}_i}{dt^2}&=&\frac{{\bf F}_i(..., {\bf r}_j, ..., {\bf v}_j, ..., {\bf p})}{m_i}\label{eq:F=mA}
\end{eqnarray}
where $m_i$ is the mass of the considered object, ${\bf r}_j, {\bf v}_j$ denote the state vectors of any body influencing the system and ${\bf p}$ vector denotes a set of any physical parameters 
relevant in the dynamics (masses, spherical harmonic coefficients $C_{np},\ S_{np}$, tidal parameters, etc.). The integration of this system consists in $3N$ differential equations and is, in most 
cases, not problematic. In the case of natural satellites, initial conditions associated with eq.(\ref{eq:F=mA}) are generally borrowed from a former ephemeris. If none is available, a simplified dynamical model may be used in a first step, with possible constrains on initial inclination and eccentricity. In the case of spacecraft, extrapolation of a former orbit (sometimes associated with an earlier phase of the mission)  may be used.

Observations useful for constraining the orbits can have very different forms: astrometric images, laser ranging, photometric, radiometric and VLBI measurements. The gathering of these 
astrometric observations is a lengthy and thankless task. Fortunately astrometric databases exist like the Planetary Data System (https://pds.nasa.gov) for space probes data and the Natural Satellite Data Base (Arlot \& Emelyanov, 2009) for the natural moons. Still, a huge amount of observations have been performed worldwide, and taking care of all various format, observation corrections, and even sometimes typos becomes extremely fastidious.

The last step requires to compare observed and computed positions of the celestial objects. In practice, one does not observe directly 3D cartesian coordinates, and the numerical output of the integration of eq. (\ref{eq:F=mA}) has to be rewritten, for each observation time, introducing observed variables like angles on the celestial sphere, ranging, Doppler, etc. These observation quantities can be dependent on the state vectors of the observed body, but also on a set of parameters ${\bf p}^\prime$ related to the observation treatment. Denoting $g$ such observation kind, and in the vicinity of the exact solution, we may express the differences between the observed and computed quantities, as a Taylor expansion, limiting ourselves to the first order
\begin{eqnarray}
g({\bf r}^o_i, {\bf v}^o_i, {\bf p}^{\prime o})-g({\bf r}^c_i, {\bf v}^c_i, {\bf p}^{\prime c})&\simeq&\sum_{l=1}^{6N+p+p^\prime}\left(\frac{\partial g}{\partial {\bf r}_i^c}\cdot\frac{\partial {\bf r}_i^c}{\partial c_l}+
\frac{\partial g}{\partial {\bf v}_i^c}\cdot\frac{\partial {\bf v}_i^c}{\partial c_l}+
\frac{\partial g}{\partial {\bf p}^{\prime c}}\cdot\frac{\partial{\bf p}^{\prime c}}{\partial c_l}\right)\Delta c_l\label{eq:O-C}
\end{eqnarray}
where $o$ and $c$ refer to observed and computed quantities, respectively, and $c_l$ denotes any unknown scalar to be fitted. Clearly, there will be as many linear equations as observation data. The linear system may then be solved by more or less sophisticated least squares method. In particular, the weight of each data, the choice of the physical parameters to be fitted  and the way these parameters may be fitted  (all in once, or using successive steps) will depend on the ephemeris developer and expertise.

In the former equation, the partial derivative of state vectors as function of initial parameters have to be known. There exists few different methods to obtain these quantities. But the most widely used 
method consists in integrating the so-called variational equations (Peters 1981; Moyer 2003). Starting from eq. (\ref{eq:F=mA}) and assuming $c_l$ to be independent of time, one obtains after applying partial derivation
\begin{eqnarray}
\frac{d^2}{dt^2} \left(\frac{\partial {\bf r}_i}{\partial c_l}\right)&=&\frac{1}{m_i} \sum_{j=1}^{N}\left[\frac{\partial {\bf F}_i}{\partial{\bf r}_j}\cdot\frac{\partial {\bf r}_j}{\partial c_l}+\frac{\partial{\bf F}_i}{\partial{\bf v}_j}\cdot \frac{d}{dt} \left( \frac{\partial {\bf r}_j}{\partial c_l} \right)\right]+\overline{\frac{\partial {\bf F}_i}{\partial c_l}}\label{eq:EQV}
\end{eqnarray}
where the last term denotes the derivation of the force with respect to $c_l$, when it comes explicitly in the expression of ${\bf F}$. The numerical integration of eq. (\ref{eq:EQV}) is much more complex than the standard equations of motions and often implies the simultaneous integration of thousands of ODE. In practice, system of eq.(\ref{eq:EQV}) needs to be integrated simultaneously with system of eq.(\ref{eq:F=mA}). This method is the one used regularly by space geodesy codes like DPODP (Moyer 1971) and GEODYN II (Pavlis et al. 2013). It is also extremely used for natural moons (Peters 1981), asteroids, comets and even for exoplanet systems. Moreover, this method can be used indifferently for regular satellite orbits, flyby analysis and even rotation analysis (replacing the equations of motion by the Euler-Liouville equations).

Last but not least, let us emphasize that the validity of eq. (\ref{eq:O-C}) implicitly assumes that the modeling is perfect and the data contains no errors. In that respect, the least squares method handles the observation errors properly if the data are not biased, only. And it is optimal if the random errors have a Gaussian profile. But a perfect modeling and no observations biases (a requirement for a Gaussian error distribution) are two conditions that are never completely reached. Combined with correlations between fitted parameters, the former method will provide different solutions, depending on the modeling, the biases treatment, the weight of the data and the selection of the released parameters.

\section{Tidal parameters quantified today}\label{sec:Tidal}

We give in Table \ref{Table:tide} the most up-to-date values of the principal tidal parameters that have been determined from astrometric measurements. When available, error bars computation 
are given, most of the time as function of the formal standard deviation. When a purely Gaussian noise is assumed, there are about 66\% of chances that the real physical value lies in the $1 \sigma$ formal uncertainty. However, as already mentioned in Section 2, observations may not follow a Gaussian error profile. More importantly, constrains are often introduced in the least squares inversion. This is generally done when working with artificial objects, since on one hand, some perturbations cannot be perfectly modeled (wheel-off loadings, orientation of solar panel, drag), and on the other hand the observables  quantities (Doppler, ranging) do not allow to recover easily the full motion of the spacecraft in 3D space. As a result, an uncertainty of $10 \sigma$ shall not be considered as necessarily more robust than a $2 \sigma$ error bar. More likely, such larger choice of error bar is a consequence of strong correlations and associated constrains inserted in the fitting process. In practice, it may be wiser to consider the error bars as the likely range in which the real physical value is lying. Sometimes, studies provide a double-check of their results by performing two independent fits (using different codes and weight) like in Iess et al. (2012). Alternatively, both independent solutions may be merged in one single solution (including error bars). Such solutions are indicated by {\it MS} (merged solutions) in Table 1.

\begin{table*}
\caption{Current estimate of the principal tidal parameters of the objects of the Solar system, as determined from astrometric measurements.}\label{Table:tide}
\begin{center}
{\footnotesize
\begin{tabular}{lcccc}
\hline
\hline
Body&$k_{lm}$&$Q$& $k_2/Q$ & Reference \\
\hline
&&&&\\
Mercury & $k_2=0.451 \pm 0.014\ (10\sigma)$   &    & & Mazarico et al. (2014)\\
&&&&\\
Venus &  $k_2=0.295 \pm 0.066 \ (2\sigma)$  &    &  & Konopliv et al. (1996) \\
&&&&\\
Earth &    &    $280\ (230, 360)\ (1\sigma)$ & & Lunar semi-diurnal terrestrial tide (Ray et al. 2001)\\ 
&&&&using $k_2=0.302$ (Wahr 1981)\\
&&&&\\
&$k_{20}=0.29525$&&&Nominal values of solid Earth tide\\
&$k_{21}=0.29470$&&&(elastic Earth) from IERS 2010\\
&$k_{22}=0.29801$&&&\\
&$k_{30}=0.093$&&&\\
&$k_{31}=0.093$&&&\\
&$k_{32}=0.093$&&&\\
&$k_{33}=0.094$&&&\\
&&&&\\
Moon  &  & $37.5 \pm 4$ & & at month frequency (Williams et al. 2014)\\
&&&& using $k_2=0.024059$ (Konopliv et al. 2013)\\
&&&&\\
&$k_{20}=0.02408 \pm 0.00045$&&&Konopliv et al. (2013) \\
&$k_{21}=0.02414 \pm 0.00025$&&&\\
&$k_{22}=0.02394 \pm 0.00028$&&&\\
&$k_2=0.02405 \pm 0.000176$&&&\\
&$k_3=0.0089 \pm 0.0021$&&&\\
&&&&\\
&$k_{20}=0.024165 \pm 0.00228$&&&Lemoine et al. (2013)\\
&$k_{21}=0.023915 \pm 0.00033$&&&\\
&$k_{22}=0.024852 \pm 0.00042$&&&\\
&$k_2=0.02427 \pm 0.00026$&&&\\
&$k_{30}=0.00734 \pm 0.00375$&&&\\
&&&&\\
Mars & $k_2=0.173 \pm 0.009\ (5\sigma)$   & $99.5 \pm 4.9$&  &Konopliv et al. (2011), Jacobson \& Lainey (2014)\\
 & $k_2=0.164 \pm 0.009\ (5\sigma)$  &  &  & after correction for atmospheric tide \\
&&&&\\
Jupiter &    &    & $(1.102 \pm 0.203) \times 10^{-5}\ (1\sigma)$&Lainey et al. (2009)\\
&&&&\\
Io &    &    &  $0.015 \pm 0.003\ (1\sigma)$ & Lainey et al. (2009)\\
&&&&\\
Saturn &  $k_2=0.390 \pm 0.024\ (MS)$  &   & $(1.59 \pm 0.74) \times 10^{-4}\ (MS)$ &Lainey et al. (2015) \\
&&&$(123.94 \pm 17.27) \times 10^{-5}\ (MS)$ & at Rhea's tidal frequency\\
&&&&\\
Titan &$k_2=0.589 \pm 0.150\ (2\sigma)$&    &  & Iess et al. (2012)\\
 &$k_2=0.637 \pm 0.224\ (2\sigma)$&    &  &   \\
&&&&\\ 
\hline
\end{tabular} 
}
\end{center} 
\label{k2Q.tab} 
\end{table*} 

Since Mercury and Venus do not have moons, the measurement of their geophysical parameters must rely on spacecraft data. Considering the semi-diurnal tides raised by the Sun, Mazarico et al. (2014) and Konopliv et al. (1996) succeeded in estimating the $k_2$ of Mercury and Venus, respectively. While Messenger's second extended mission data might provide a better determination in the future, no estimation of Venus' $k_2$ has been published since 1996, and the analysis of the Pioneer Venus Orbiter and Magellan space probes. In particular, Venus Express orbit is too far and eccentric to provide a good S/N ratio on Venus' $k_2$.

On the contrary, Mars, Jupiter and Saturn's system have moons, close enough to provide reliable signal over a century of accurate astrometric observations. In particular, it is astonishing that the secular acceleration of Phobos around Mars was first pointed out in 1945 by Sharpless (1945) with a 50\% error in his determination, only. While the acceleration of Phobos provides the tidal ratio $k_2/Q$, it is the Mars' spacecraft that allow to quantify the Mars $k_2$ associated to Solar tides. Assuming the same Mars'  $k_2$ at Solar and Phobos tidal frequency, Jacobson \& Lainey (2014) published the most up-to-date $Q$ value.

While the first estimation of the Jovian $k_2$ is a goal of the Juno mission, Lainey et al. (2009) succeeded in fitting the tidal ratio $k_2/Q$ of both Jupiter and Io, using astrometric measurements of the Galilean moons. In a similar way, Lainey et al. (2015) used a large set of astrometric data, including numerous ISS-Cassini data to quantify Saturn's $k_2/Q$ and its dependence on tidal frequency. More, they used astrometric observations of the Lagrangian moons of Tethys and Dione to provide the first determination of the Saturn's $k_2$. Still with Cassini spacecraft, but this time from the radiometric measurement of six flybys of Titan, Iess et al. (2012) could provide the first determination of Titan's $k_2$, suggesting a potential global ocean under the moon's shell.

The Earth-Moon system is evidently the most studied, but also the most complex. In particular, the presence of oceans, atmosphere and convection in the mantle of the Earth makes the data treatment quite difficult. For the Moon, the latest Love numbers estimations were obtained from the GRAIL mission, with two independent published studies (Konopliv et al. 2013, Lemoine et al. 2013). Benefiting from GRAIL mission results, Williams et al. (2014) reanalyzed lunar laser ranging data and provided the latest estimation of the Lunar tidal quality factor $Q$ at month frequency.

The difficult proper treatment of Earth data and modeling made the determination of terrestrial tides extremely difficult. Its first determination arose in 1996 only (Ray et al. 1996), and was improved in 2001 (Ray et al. 2001). To conclude this section, we provide in Table 1 the nominal values of Love numbers for a solid (elastic) Earth tide from IERS 2010. The interested reader may find more informations on Earth's tides in IERS (2010) and references therein.

\section{Possible determination of tidal parameters for the icy giant planets of our Solar system}\label{sec:IcyG}

\begin{figure*} 
\begin{center} 
\includegraphics[width=12.cm,angle=0]{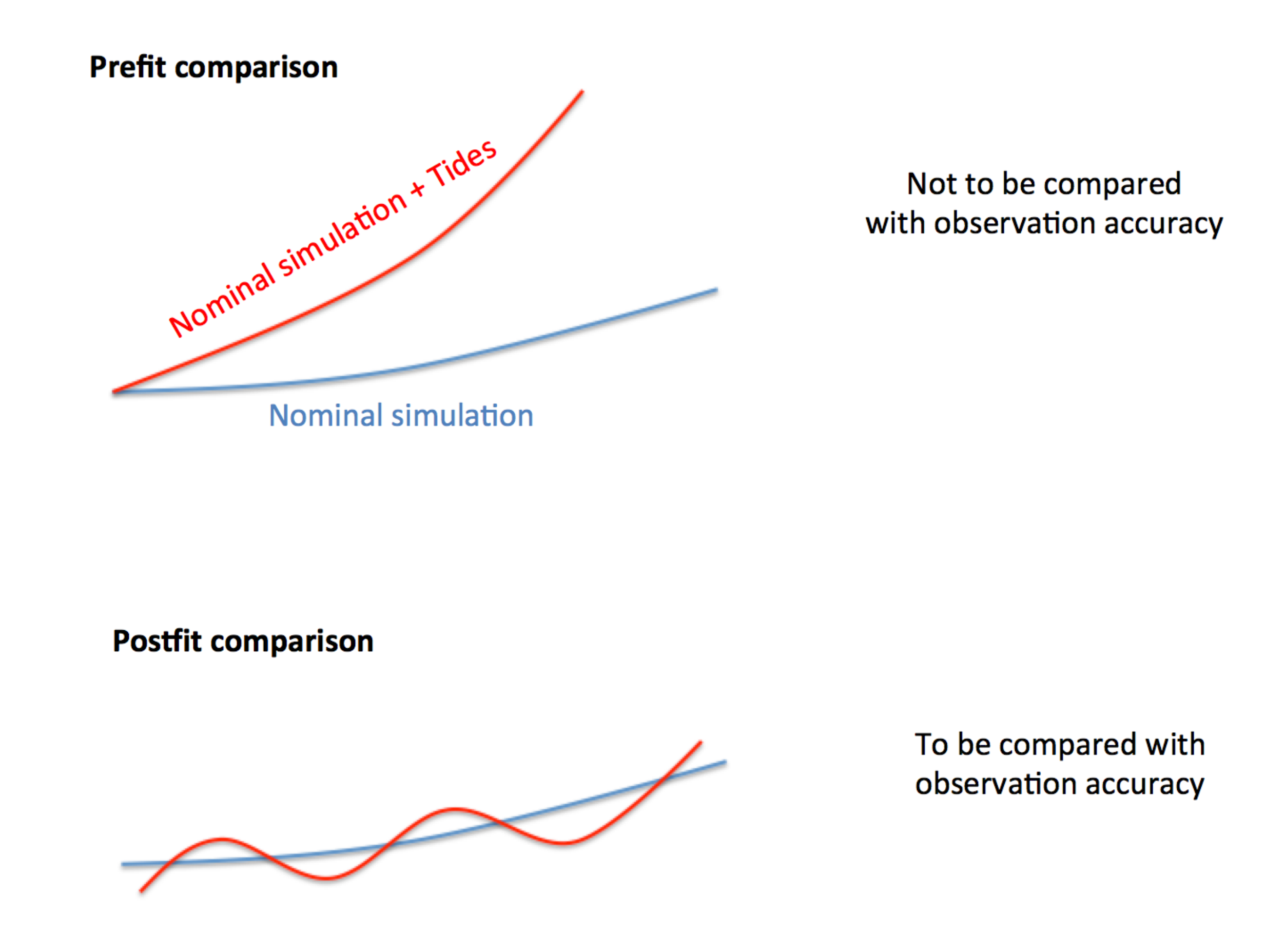}
\caption{The retained method for checking the detectability of tides.}\label{fig:schemeFit}
\end{center} 
\end{figure*}

To investigate the possible detection of secular acceleration of moons associated with tidal dissipation within Uranus and Neptune, we rely on numerical integration.
A simple look at the differences on the positions of the moons after adding/removing the tidal effects over 100 years (roughly the time span of accurate observations) is
meaningless. Indeed, one needs to take into account the fitting procedure of the initial conditions to the observations (see Section \ref{methodo}). In particular, 
the difference in modelling may be partly masked in a slight change of the initial conditions (Lainey \& Tobie 2005). As a consequence, the true incompressible part of the tidal effects in the dynamics is revealed only after having fitted one simulation onto the other (see Figure \ref{fig:schemeFit}). Detection threshold for a perturbation is highly dependent on the error profile (Gaussian noise may easily get averaged) and observation number $n$. For a Gaussian error profile, the uncertainty on the value of the fitted parameters is linearly dependent on the standard deviation $\sigma$ of the observation error and inversely proportional on $\sqrt{n}$. Here, considering that post-fit biases represent typically between few to few tens of percents of the global astrometric residuals (Lainey et al. 2009), we will consider heuristically that a physical signal may be fitted, at the very extreme, up to a level of a tenth of the astrometric residuals.

The fitting residuals are given in right ascensions (RA) and declinations (DEC) variables, much closer to real observation data. A difference of a factor of two was applied in the weight between observations performed during the first and last 50 years, to somewhat mimic the difference in accuracy between the old and modern observations. For both systems, we tested three different cases: $Q=10,000, 1000, $ and $100$. The Love numbers $k_2$ were borrowed from Gavrilov \& Zharkov (1977) and set to $k_2=0.104$ and $k_2=0.127$ for Uranus and Neptune, respectively.
Numerical simulations were performed over +/- 50 years to limit numerical errors. All residuals showed a linear dependency to the ratio $k_2/Q$, as expected from the linearity of the analytical evolution of the semi-major axes (Goldreich \& Soter 1966). Here, we provide post-fit residuals for the extreme case $Q=100$, only. Results for the $Q=1000$ and $Q=10,000$ solutions can easily be inferred by applying a rescaling coefficient on the residuals of a factor of 10 ($Q=1000$) and 100 ($Q=10,000$), respectively.

\subsection{The Uranus system}

\begin{figure*} 
\begin{center} 
\begin{tabular}{l} 
\includegraphics[width=8.cm,angle=-90]{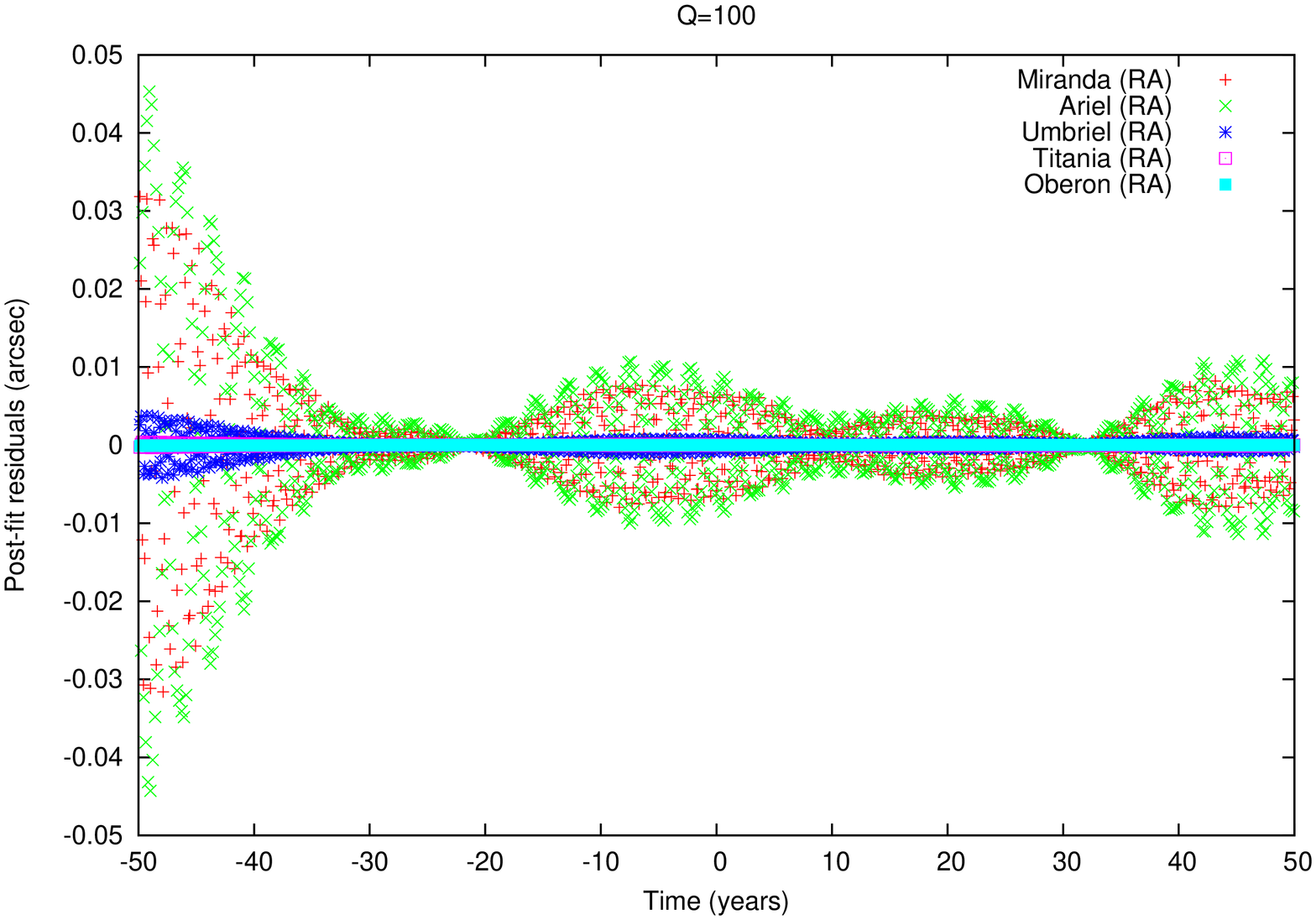}\\
\includegraphics[width=8.cm,angle=-90]{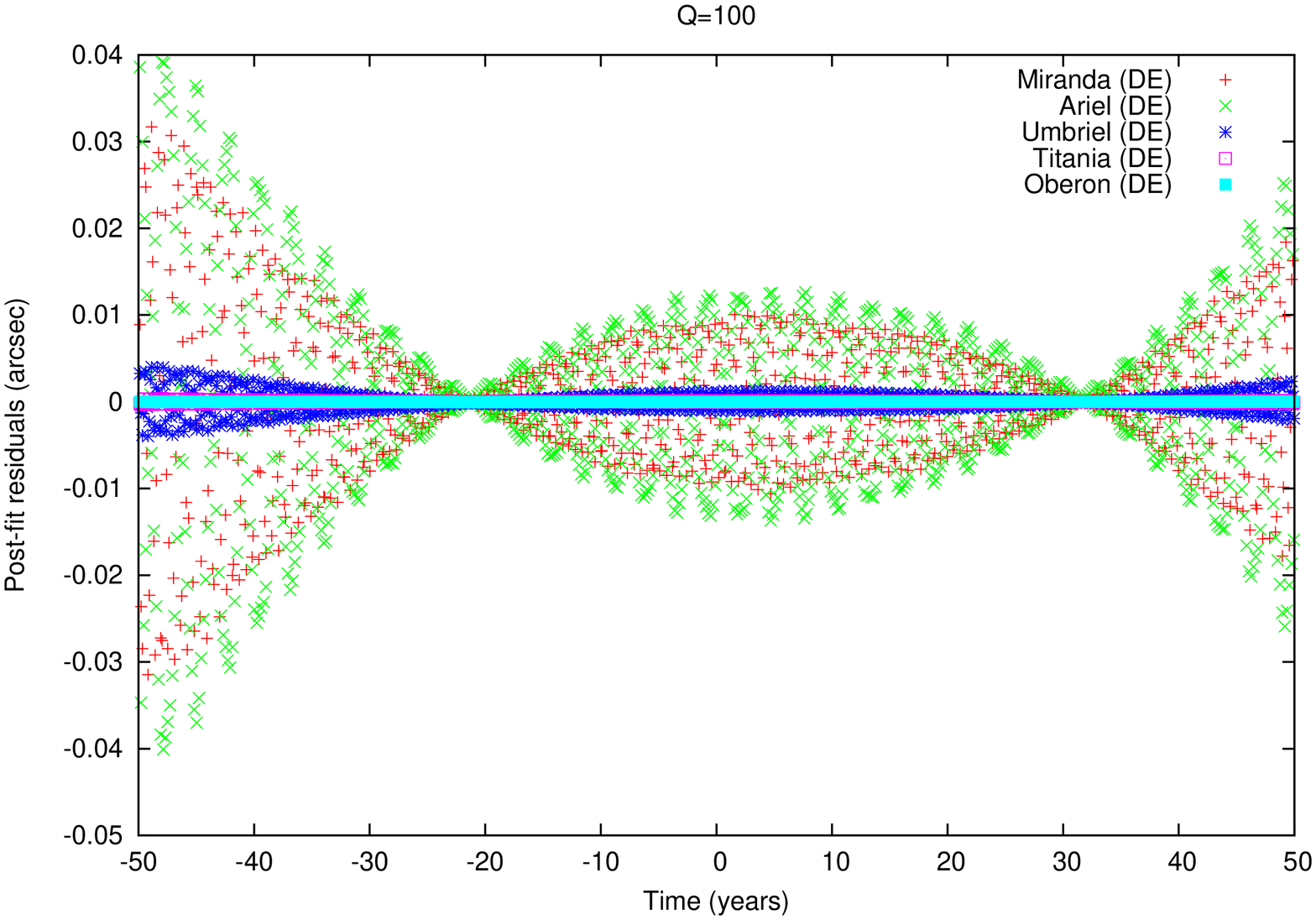}
\end{tabular} 
\caption{Differences associated to tides on right ascension (top) and declination (bottom) of the five main Uranian moons. We recall that 0.01 arcsecond is about 130 $km$ at the opposition.}\label{fig:URA}
\end{center} 
\end{figure*}

Following Lainey (2008), we considered in the simulations the five main moons and the Sun as interacting bodies, only. The modeling and initial conditions of the system were pretty close 
to the ones used by Lainey (2008). The residuals for the $Q=100$ case are given in Figure \ref{fig:URA} and reach up to 50 mas on RA and 40 mas on DE on Ariel's orbit. This translates into about 600 km and 500 km at opposition, respectively. Such differences are large compared to the Voyager astrometric data (few tens of kilometers accuracy) or even the first mutual event campaign of the Uranian moons 
(few tens of mas accuracy). If these data provide a strong constrain on a very short period of time, constraining the tidal accelerations requires accurate data dispatched over a long time interval. 
In that case, the observation sets of the Uranian moons have a typical accuracy of 60 mas to 300 mas (Veiga et al. 2003, and Tables 1-3 in Lainey 2008). Because of its faint brightness and closer distance to the rings,
Miranda has generally much larger residuals, and may not be useful for constraining the Uranian tidal dissipation. All in all, Ariel's orbit seems to be the best candidate, from far, for constraining 
Uranian tides. Considering the possibility to fit a dynamical parameter that provides an incompressible signal at the level of one tenth of the amplitude of the astrometric residuals, we deduce that a 10 mas signal should be detectable from astrometric observations. This puts a higher limit of $Q=500$ for being able to fit Uranian tidal dissipation (assuming $k_2=0.104$ from 
Gavrilov \& Zharkov (1977)). Such limit may be relaxed with the new reduction of old observations, like what was done recently for the Mars and Jupiter system (Robert et al. 2015; 2011).

Concerning new observation campaign, the regular astrometric campaigns from ground may hardly obtain a better accuracy than few tens of mas. 
However, the Gaia mission shall provide about 50 astrometric observations of the Uranian moons, except Miranda too close to its primary, with an accuracy at the mas level. Such accuracy is close to 
the accuracy of the Voyager 2 data and will be extremely valuable for constraining tides. As a consequence, ground support may consider focusing on fainter Uranian moons, starting with Miranda itself. 

\subsection{The Neptune system}

\begin{figure*} 
\begin{center} 
\begin{tabular}{l} 
\includegraphics[width=8.cm,angle=-90]{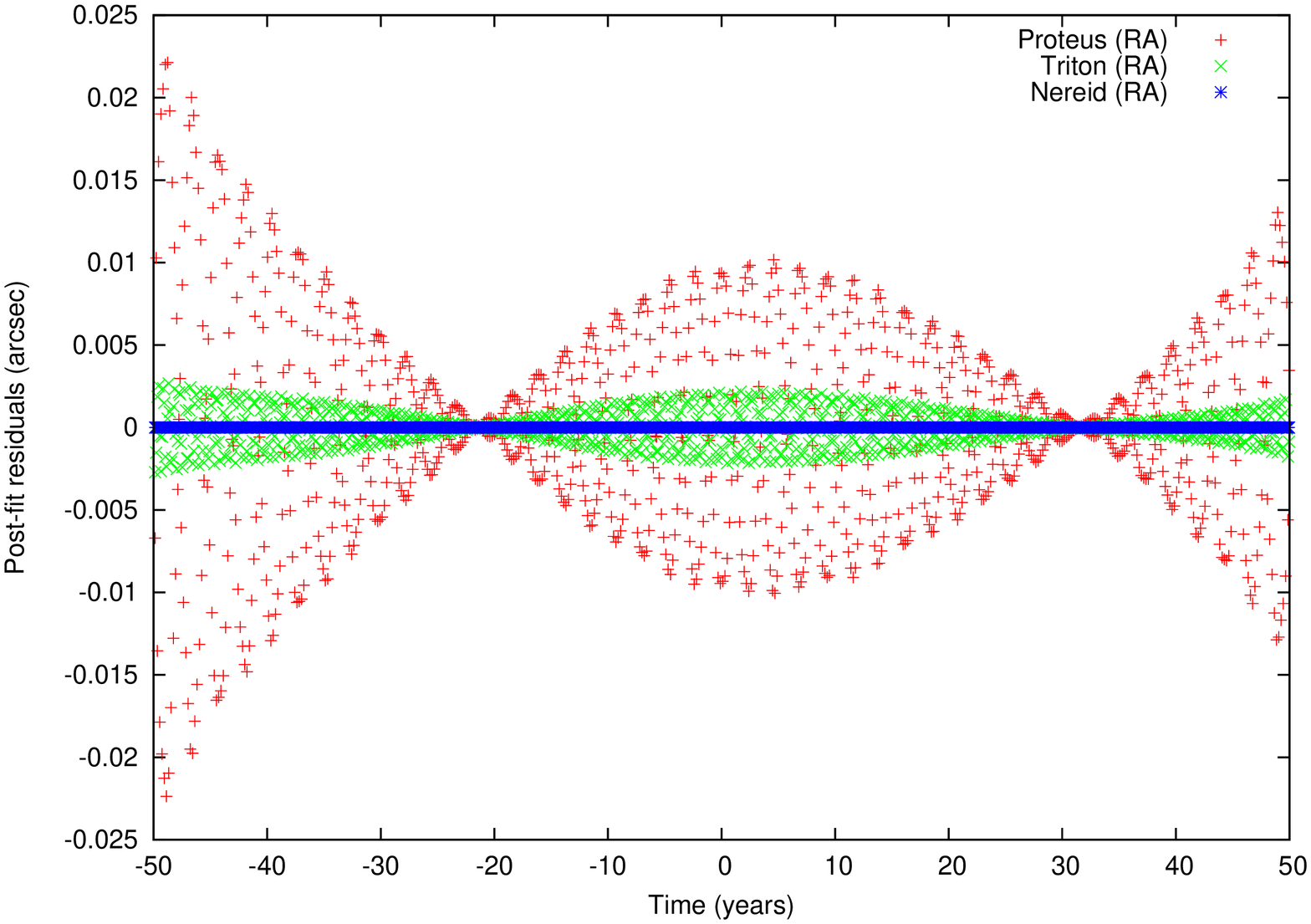}\\
\includegraphics[width=8.cm,angle=-90]{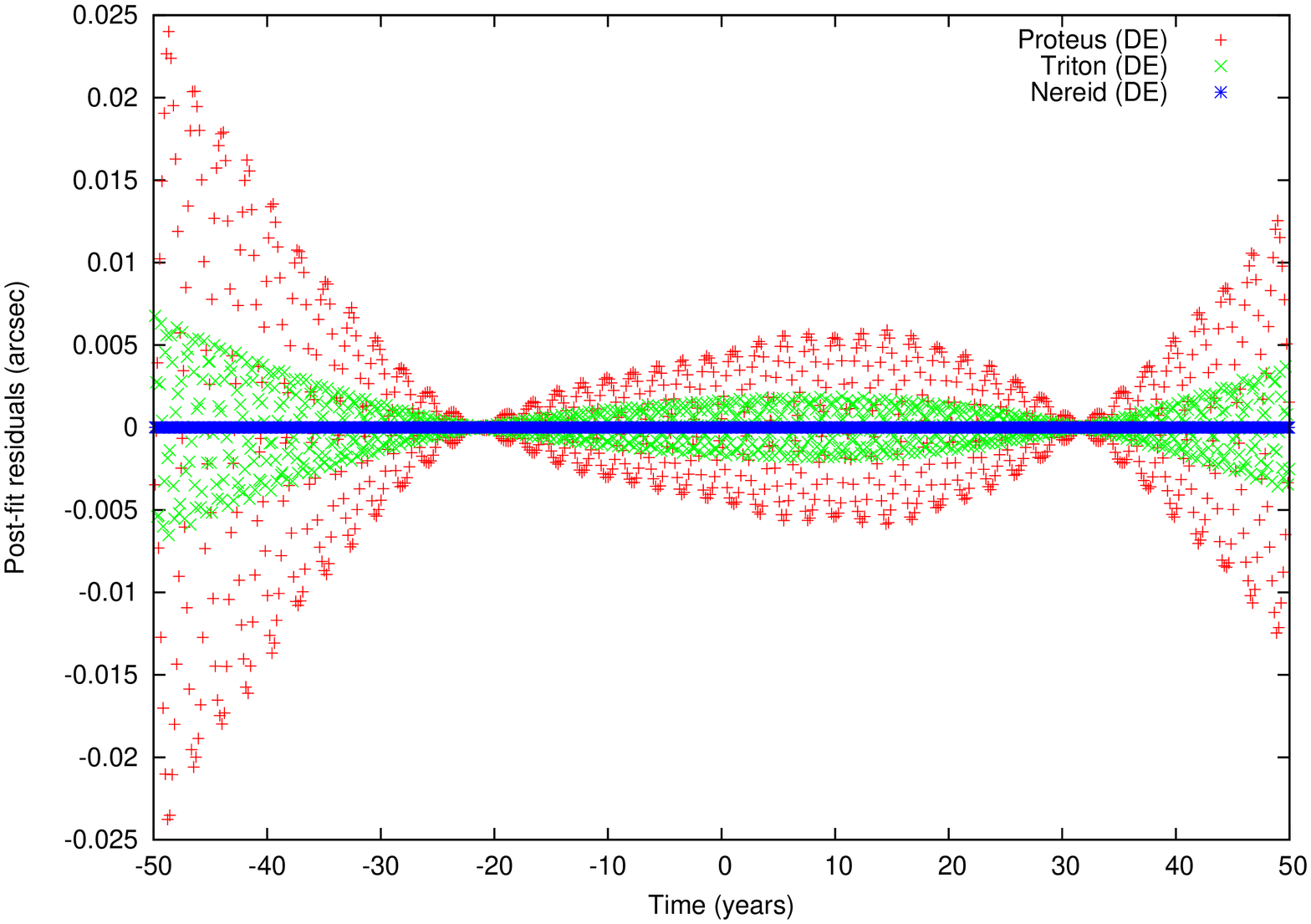}
\end{tabular} 
\caption{Differences associated to tides on right ascension (top) and declination (bottom) of the three Neptunian moons, Proteus, Triton and Nereid. We recall that 0.01 arcsecond is about 210 $km$ at the opposition.}\label{fig:NEP}
\end{center} 
\end{figure*}

In these simulations we considered the interaction of Proteus, Triton, Nereid, the Sun and Uranus. In a first step, our initial conditions were fitted to the SPICE kernel nep081 (Jacobson, 2009) to allow for realistic orbital simulations. The residuals for the $Q=100$ case are given in Figure \ref{fig:NEP} and reach up to 25 mas on both RA and DE on Proteus' orbit. This translates into about 525 $km$ at 
opposition. Unfortunately, this moon was discovered in 1989 from Voyager 2 images. As a consequence, the real available data span for Proteus is roughly a third of a century, only. Since tidal effects induce a 
quadratic behaviour on the moons' longitudes, the magnitude of Neptune's tides on Proteus' orbit should be reduced by a factor $3^2$, that is an order of magnitude. This is about a few mas, 
at the same level of Triton's residuals on Figure \ref{fig:NEP}. Such signal seems too small to be fitted from astrometric data.

\section{Conclusion}

Even though of high importance for constraining interior properties and the dynamical evolution of celestial objects, tidal parameters remain difficult to determine. We have recalled briefly the fundamental method used to characterize the tidal parameters directly from observations. It appears that about ten of Solar system objects have tidal parameters being fitted, only. Simulations performed over a century of 
the orbit of the most observed Uranian and Neptunian moons suggest that the determination of tidal parameters will be extremely difficult, unless the ratio $k_2/Q$ is pretty high. In that context, a space mission dedicated to icy giant planets should provide quite valuable data, and even more accurate than the observations of Gaia mission. The consideration of all (past, present, and possibly future) space mission data, in addition to ground data (possibly rereduced with Gaia's star catalog) shall allow for a determination of the tidal dissipation (if significant) in Uranus.


\begin{acknowledgements}
V.L. is grateful to P.Rosenblatt, J.C.Marty and B.Noyelles for fruitful discussions.  
This work has been supported by the International Space Science Institute (ISSI), the scientific council of the Paris Observatory and the PNP (INSU/CNES).
\end{acknowledgements}



\begin{thebibliography}{}
%

%


\bibitem{Archinal et al. (2011)} Archinal, B.A., et al.:
Report of the IAU Working Group on Cartographic Coordinates and Rotational Elements: 2009
Celestial Mechanics and Dynamical Astronomy {\bf 109}, 101-135 (2011)

\bibitem{Arlot & Emelyanov (2009)} Arlot, J.-E. and Emelyanov, N.V.:
The NSDB natural satellites astrometric database
Astronomy and Astrophysics {\bf 503}, 631-638 (2009)

\bibitem{Crida & Charnoz (2012)} Crida, A. and Charnoz, S.:
Formation of Regular Satellites from Ancient Massive Rings in the Solar System
Science {\bf 338}, 1196- (2012)

\bibitem{?uk (2014)} \'Cuk, M.: 
Recent Origin of Titan's Orbital Eccentricity. American Astronomical
Society, DDA meeting 45, 301.01 (2014).

\bibitem{Gavrilov & Zharkov (1977)} Gavrilov, S.V. and Zharkov, V.N.:
Love numbers of the giant planets
Icarus {\bf 32}, 443-449 (1977)

\bibitem{Goldreich & Soter (1966)} Goldreich, P. and Soter, S.:
Q in the Solar System
Icarus {\bf 5}, 375-389 (1966)

\bibitem{IERS 2010} Petit, G., and Luzum, B.:
IERS conventions (2010)
IERS Technical Note No. 36 (2010)

\bibitem{Iess et al.(2012)} Iess, L., et al.:
The Tides of Titan
Science {\bf 337}, 457-459 (2012)

\bibitem{Konopliv et al. (2013)} Konopliv, A.~S., et al.:
The JPL lunar gravity field to spherical harmonic degree 660 from the GRAIL Primary Mission
Journal of Geophysical Research (Planets) {\bf 118}, 1415-1434 (2013) 

\bibitem{Konopliv et al. (2011)} Konopliv, A.~S., et al.:
Mars high resolution gravity fields from MRO, Mars seasonal gravity, and other dynamical parameters
Icarus {\bf 211}, 401-428 (2011)

\bibitem{Konopliv & Yoder (1996)} Konopliv, A.~S. and Yoder, C.~F.:
Venusian $k_2$ tidal Love number from Magellan and PVO tracking data
Geophysical Research Letter {\bf 23}, 1857-1860 (1996) 

\bibitem{Jacobson & Lainey (2014)} Jacobson, R.~A. and Lainey, V.:
Martian satellite orbits and ephemerides
Planetary and Space Science {\bf 102}, 35-44 (2014)

\bibitem{Jacobson (2009)} Jacobson, R.A.:
The Orbits of the Neptunian Satellites and the Orientation of the Pole of Neptune
The Astronomical Journal {\bf 137}, 4322-4329 (2009)

\bibitem{Lainey et al. (2015)} Lainey, V., et al.:
New constraints on Saturn's interior from Cassini astrometric data
arXiv:1510.05870 

\bibitem{Lainey et al. (2009)} Lainey, V., et al.:
Strong tidal dissipation in Io and Jupiter from astrometric observations
Nature {\bf 459}, 957-959 (2009)

\bibitem{Lainey (2008)} Lainey, V.:
A new dynamical model for the Uranian satellites
Planetary and Space Science {\bf 56}, 1766-1772 (2008)

\bibitem{Lainey & Tobie (2005)} Lainey, V. and Tobie, G.:
New constraints on Io's and Jupiter's tidal dissipation
Icarus {\bf 179}, 485-489 (2005)

\bibitem{Lemoine et al. (2013)} Lemoine, F.~G., et al.:
High-degree gravity models from GRAIL primary mission data
Journal of Geophysical Research (Planets) {\bf 118}, 1676-1698 (2013)

\bibitem{Love 1911} Love, A. E. H.: 
Some Problems of Geodynamics (Cambridge University Press) (1911)

\bibitem{Matson et al. 1981} Matson, D. L., et al.:
Heat flow from Io 
Journal of Geophysical Research {\bf 86}, 1664-1672 (1981)

\bibitem{Mazarico et al. (2014)} Mazarico, E., et al.:
The gravity field, orientation, and ephemeris of Mercury from MESSENGER observations after three years in orbit
Journal of Geophysical Research (Planets), 119, 2417 (2014)

\bibitem{Moyer (2003)} Moyer, T.D.:
Formulation for observed and computed values of Deep Space Network observables
576pp., Wiley, Hoboken, N.J. (2003)

\bibitem{Moyer (1971)} Moyer, T.D.:
Mathematical formulation of the double-precision orbit determination program (DPDOP)
Technical report 32-1527, National Aeronautics and Space Administration (1971)

\bibitem{Pavlis et al. (2013)} Pavlis, D.E., et al.:
GEODYN II system description, vol. 1-5, contractor report, SGT Inc., Greenbelt, Md. (2013)

\bibitem{Peters (1981)} Peters, C.~F.:
Numerical integration of the satellites of the outer planets
Astronomy and Astrophysics {\bf 104}, 37-41 (1981)

\bibitem{Ray et al. (2001)} Ray, R.~D., et al.:
Constraints on energy dissipation in the Earth's body tide from satellite tracking and altimetry
Geophysical Journal International {\bf 144}, 471-480 (2001)

\bibitem{Ray et al. (1996)} Ray, R.D., et al.:
Detection of tidal dissipation in the solid Earth by satellite tracking and altimetry
Nature {\bf 381}, 595-597 (1996)

\bibitem{Robert et al. (2015)} Robert, V., et al.:
A new astrometric measurement and reduction of USNO photographic observations of Phobos and Deimos: 1967-1997
Astronomy and Astrophysics {\bf 582}, A36 (2015)

\bibitem{Robert et al. (2011)} Robert, V., et al.:
A new astrometric reduction of photographic plates using the DAMIAN digitizer: improving the dynamics of the Jovian system
Monthly Notices of the Royal Astronomical Society {\bf 415}, 701-708 (2011)

\bibitem{Sharpless 1945} Sharpless, B.P.:
Secular accelerations in the longitudes of the satellites of Mars
Astronomical Journal {\bf 51}, 185-186 (1945)

\bibitem{Spencer et al. 2006} Spencer, J.R., et al.: 
Cassini encounters Enceladus: Background and the discovery of a south polar Hot Spot
Science {\bf 311}, 1401-1405 (2006)

\bibitem{Veiga et al. (2003)} Veiga, C.H., et al.:
Positions of Uranus and Its Main Satellites
Astron. J. {\bf 125}, 2714-2720 (2003)

\bibitem{Wahr (1981)} Wahr, J.M.:
Body tides on an elliptical, rotating, elastic and oceanless earth
Geophys. J. R. astr. Soc. {\bf 64}, 677-703 (1981)

\bibitem{Williams et al. (2014)} Williams, J.~G., et al.:
Lunar interior properties from the GRAIL mission
Journal of Geophysical Research (Planets) {\bf 119}, 1546-1578 (2014)


\end{thebibliography}
\end{document}